# Machine learned potential for defected single layer hexagonal boron nitride

John Janisch,[1] Duy Le,[1,2] and T.S. Rahman[1,2,3]

[1]Department of Physics, University of Central Florida, Orlando, Florida 32816, United States

[2]Renewable Energy and Chemical Transformations (REACT) Cluster, University of Central Florida, Orlando, Florida 32816, United States

3Donostia International Physics Center, Paseo de Manuel Lardizábal 4, 20018, San Sebastian, Spain.

**Abstract:**

Development of machine learned interatomic potentials (MLIP) is critical for performing reliable simulations of materials at length and time scales that are comparable to those in the laboratory. We present here a MLIP suitable for simulations of the temperature dependent structure and dynamics of single layer hexagonal boron nitride (*h*-BN) with defects and grain boundaries, developed using a strictly local equivariant deep neural network as formulated in the Allegro code. The training dataset consisted of about 30,000 images of *h*-BN with and without point defects generated with *ab-initio* molecular dynamics simulations, based on density functional theory (DFT), at 500, 1000, and 1500°K. The developed MLIP predicts potential energies and forces with a mean absolute error (MAE) of 4 meV/atom and 60 meV/Å, respectively. It also reproduces phonon dispersion curves and density of vibrational states of pristine bulk *h*-BN that are comparable with that obtained from density functional theory-based calculations. Molecular dynamics simulations of the motion of the 4|8 grain boundary unit in *h*-BN shows the first step to have an activation barrier ~2.2 eV, indicating immobility of the grain boundary. Moving the grain boundary units past the first shows much lower activation barriers of ~0.42eV, suggesting a facile motion of the grain boundary once the first movement is stimulated. These simulations yield a scaled mobility of $1.739 \times 10^{-11} \frac{m^3}{Js}$ for a temperature of 1500°K which, given the inherent differences in the set-ups, is not too far from the experimental value of $1.36 * 10^{-9} \frac{m^3}{Js}$. The ability to predict grain boundary mobility within reasonable agreement with experiment demonstrates the robustness of the MLIP and its suitability for reliable simulations of defect structures and dynamics in single layer *h*-BN.

# 1 Introduction

Hexagonal boron nitride (*h*-BN) is an interesting material for a variety of reasons. Its high thermal conductivity, usefulness as a nanoelectronics component, and its anti-corrosion properties make it an attractive material for electronic applications [1-5]. It is also utilized for medical purposes thanks to its chemical inertness [6,7]. Additionally, *h*-BN is shown to be an excellent choice for mechanical applications since single-layer *h*-BN is one of the strongest electrical insulators and adding layers of *h*-BN does not reduce the mechanical strength of the multi-layer structure [8]. These applications represent only a fraction of *h*-BN's potential, as many more possibilities emerge when considering defects. Inherent defects, such as vacancies, have shown to render *h*-BN a single photon emitter in the mid-infrared range near room temperature [9,10]. Depending on the size and termination of vacancies and nanoholes, magnetic properties can arise in the otherwise non-magnetic *h*-BN sheets [11]. On the other hand, when *h*-BN sheets are doped with Li they can be useful for hydrogen storage [12] and when metal atoms are introduced to a pristine *h*-BN sheet the resulting system can be catalytically active for reactions including CO oxidation [13,14]. Defected *h*-BN has also been shown to be an effective metal-free catalyst for the hydrogenation of olefins [15] as well as being shown computationally and experimentally to be a potential catalyst for capturing $CO_2$ and converting it to methanol and formic acid [16,17].

Our interest in defect laden *h*-BN, because of some of the novel functionalities mentioned above, dictates the need for a comprehensive understanding of the geometry and atomic scale properties of the system. While Density Functional Theory (DFT) is one of the best tools for investigating the ground state characteristics, it is computationally unrealistic to simulate the dynamics of systems consisting hundreds of thousands of atoms, as would be the case for *h*-BN sheets with defects and grain boundaries, to encompass not only the defect and its local environment, but also the long reaching effects of defect-defect interactions which would extend for several nanometers. Although an energy minimization calculation on an expansive simulation box is not unrealistic with DFT, carrying out ab initio molecular dynamics (MD) simulations for any meaningful time scale would be computationally intensive. This is where the application of machine learning comes into play.

Machine learned interatomic potentials (MLIPs) have been shown to be many orders of magnitude faster than DFT with nearly the same level of accuracy. For example, Xie et al. [18] reported a speedup of 3-6 orders of magnitude over DFT for a simulation with 128 tungsten atoms, depending on the specific MLIP architecture that was used, making the speed up with MLIPs quite apparent.

In recent years MLIP has been applied to examine characteristics of *pristine h*-BN such as stress response at various temperatures [19], thermal conductivity [20,21], crystallization under different conditions, and the mechanical and thermal properties of the resulting crystal [22] to name a few. Using a machine learned algorithm as a predictive tool involves not creating a force field but also obtaining a large dataset of simulations with a high-accuracy and having it learn patterns from that data. These predictive models take the same atomic positions as input but instead of outputting forces, energy, charges, etc. they output prediction of one or more material property of interest. One example is the work of Mousavi and Montazeri who created a machine learned model to predict the Young's modulus, the ultimate tensile strength, and fracture strain for defected *h*-BN [23]. Nevertheless, machine learned applications have their shortcomings, namely a limited applicable regime. An MLIP that is trained specifically for pristine *h*-BN will only work for pristine *h*-BN and deviating from that regime results in a failure of the MLIP. This issue is even more apparent for predictive modeling, since models trained only to predict material properties can predict only the properties they were trained to predict. Developing a robust MLIP that is applicable for a wide range of configurations of the system would allow for calculations to be performed at conditions closely matching experiments, making it easier to explain a variety of phenomena and the microscopic mechanisms responsible for them.

In this work, we will demonstrate using a dataset created from DFT calculations of pristine and single point defects of *h*-BN that we can train an MLIP that is applicable not only to the systems it was trained on, pristine and single point defects *h*-BN, but also to systems that are not part of the training dataset, i.e., grain boundaries. We will use that MLIP to study the mobility of 4|8 grain boundaries in *h*-BN which will be verified against experimental data [24] in which the motion of grain boundaries in *h*-BN was investigated. These experiments suggested that the motion of a grain boundary in *h*-BN was a concerted motion of all the atoms in the grain boundary at once, but our results will show that the individual grain boundary units d sequentially to move the grain boundary. Our simulations suggest that the initial motion of the grain boundary needs a stimulus, but once initiated the simulated grain boundary motion and calculated mobility of the grain boundary are not far from values obtained from experiments. The rest of the paper is organized as follows. In Section 1, we will discuss the choice of our specific MLIP framework. Section 2 contains a description and illustration of the dataset that was used to train our model. In Section 3, the details of the hyperparameters of our model will be given as well as the details for training settings and training results. Section 4 focuses on validation of our model through comparison with results from DFT and that already available in the literature. Finally, Section 5 will show results for our investigation of the movement of grain boundaries in *h*-BN, illustrating the applicability of our model to a wide range of systems.

## 2 Framework for development of a machine learned interatomic potential

The field of machine learned interatomic potential has been growing, resulting in several frameworks for their acquisition [25-27]. Each of them has its own pros and cons. As we discuss in the Supplementary Information, after a couple of trials with several MLIP frameworks, we chose the Allegro model [28] for training our MLIP. Allegro [28] is a strictly local equivariant deep neural network interatomic potential architecture that exhibits simultaneously excellent accuracy and scalability. It is an E3-equivariant MLIP, that is it preserves translational, rotational, and reflectional symmetry properties in 3D Euclidean space. Allegro represents the local environment around each atom using a multi-layer perception, which is another name for the feed-forward neural network shown in Figure S1 in the Supplementary Information. With the Allegro framework, there are formally only a few parameters that need to be set initially, but in practice there several customizations to be performed since the neural networks for two-body interactions, embedding, latent operations, and the final output can all be modified individually as desired. The number and size of layers, types of nonlinear functions, and initialization of weights can all be specified for each of the neural networks, meaning there are more hyperparameters that can be changed in an Allegro model than there are in a basic neural network. The factors, equivariance, local environment representation, and extensively customizable hyperparameters, influenced our choice to use Allegro.

## 3  Dataset creation

Our training dataset comprises ~30,000 images generated from *ab-initio* molecular dynamics (AIMD) simulations of a variety of 6×6 *h*-BN sheets with one of the point defects, shown in Figure 1, as well as a pristine sheet of *h*-BN. The 6 point defects, shown in Figure 1, are *h*-BN with one BN pair that is rotated by 180° (B/N) (Figure 1a), a Stone-Wales defect which takes a BN pair and rotates it by 90° (SW) (Figure 1b), a nitrogen substitution for a boron atom ($B_N$) (Figure 1c), a boron substitution for a nitrogen atom ($N_B$) (Figure 1d), and finally $V_N$ and $V_B$ are vacancies of nitrogen and boron atoms, respectively ((Figure 1e and f). Each simulation contained 72 atoms (vacancy systems contained 71 atoms) in the supercell. We use Vienna *Ab-initio* Simulation Package (VASP) [29,30] to perform DFT-based AIMD simulations employing the projector-augmented wave (PAW) [31,32] and plane wave basis set methods. We use the PAW pseudopotential, generated using Perdew-Burke-Ernzerhof functional (PBE).[33] The cut-off energy for the expansion of the plane-wave basis set is 500 eV. We sample the Brillouin Zone with one point at the zone center, and we use Gaussian smearing of 0.1 eV. For the self-consistent field (SCF) iterations we set an energy convergence threshold of $10^{-6}$ eV. Note here that in order to have a dataset that is not confined to a particular trajectory, our MD simulations are somewhat atypical. We used a time step of 2 and 5 fs and we did not thermally-equilibrate the system to create simulations that would generate a broad set of configurations that are far from equilibirum, which expands the regime in which our MLIP is applicable. We performed simulations in a microcanonical ensemble at 500, 1000, and 1500 K. In addition to the system that is constructed with the optimized lattice constant of 2.509 Å of *h*-BN, we also performed AIMD simulation with a strain of ±5% applied to the system for richness of dataset.

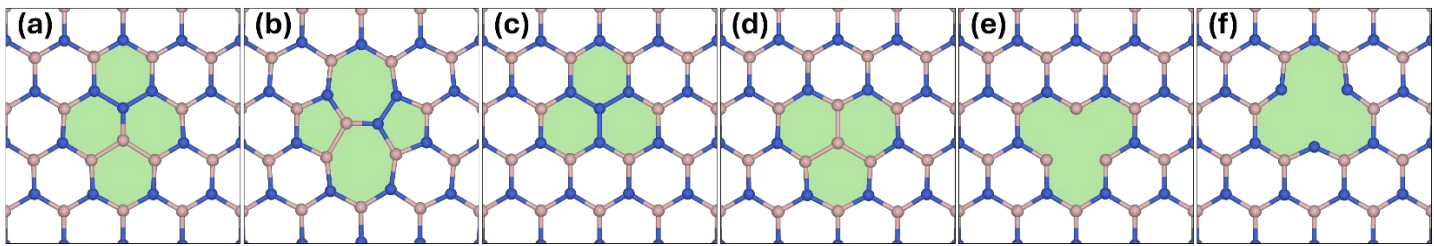

*Figure 1: Visualizations of the point defects included in the training dataset. a) 180° defect with a single BN pair rotated by 180°; b) Stone-Wales defect with a single BN pair rotated by 90°; c) $B_N$ defect with a boron atom replaced with a nitrogen atom; d) $N_B$ defect*

with a nitrogen atom replaced with a boron atom; e) $V_N$ with a vacancy created by removal of a nitrogen atom. F) $V_B$ with a vacancy created by removal of a boron atom.

Pristine systems with lattice constants of 2.45, 2.55, 2.65, and 2.75 Å were used to enhance the MLP's ability to predict structures beyond a pristine "correct" sheet of h-BN. "Correct" here means small variation from an ideal bond length between the atoms. Using entire sheets of atoms at these larger bond lengths meant a larger sample size as compared to using one with only a few atoms.

The training dataset was randomly selected from the 30,000 images and included 17,000 images used for training and 4,000 images for validation, with the data being shuffled for each training epoch. We used 2 layers as well as a cutoff of 6.0 Å, our $l_{max}$ was set to 2, and we used o3_restricted parity. The 2-body latent multi-layer perceptron (M-LP) was made up of 2 layers each with 32 nodes and utilized SiLU nonlinearities [34]. The latent M-LP had a single layer with 32 nodes and used SiLU nonlinearities as well. The embedding M-LP was a single matrix multiplication with no nonlinearity. The last edge energy M-LP was made of a single layer with 32 nodes with no nonlinearity. The basis encoding was done using a set of 8 Bessel functions with a polynomial envelope function using $p = 48$. The initial learning rate was set to 0.01 with a batch size of 50 and the learning rate was reduced according to an on-plateau scheduler using the loss function with a patience of 100 and a decay factor of 0.5, meaning if the loss function does not improve for 100 epochs the learning rate would be reduced by a factor of 0.5. The loss function was based on the total error for forces and per-atom mean square error (MSE) for energy, with a weight of 0.5 for force and a weight of 1 for energy. We used the Adam optimizer [35] as implemented in PyTorch [36] with the default parameters, so $\beta_1 = 0.9, \beta_2 = 0.999$ and $\epsilon = 10^{-8}$ without any weight decay. The training was stopped if (i) the validation loss function does not improve for 50 epochs, (ii) the validation force mean absolute error (MAE) dropped below 0.01 eV/Å, (iii) the validation energy per atom MAE dropped below 0.001 eV/atom, or (iv) if the learning rate dropped below $10^{-5}$. At the end of the training, we obtained an MAE of approximately 4 meV/atom for energy and 60 meV/Å for force (Figure 2).

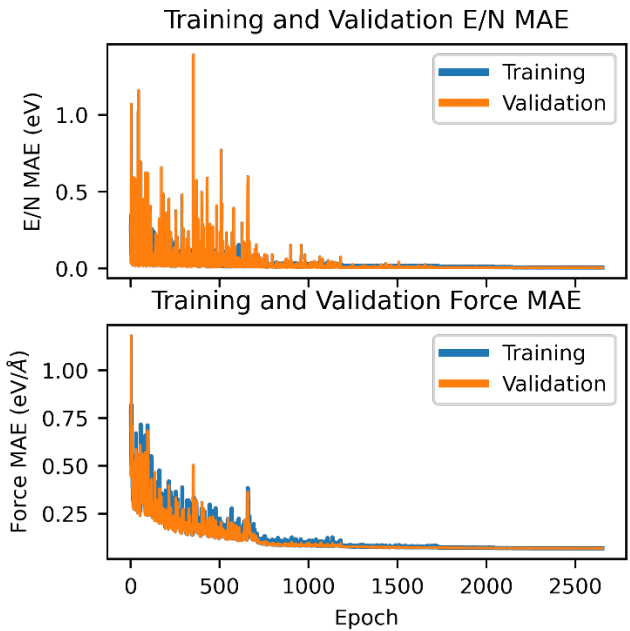

*Figure 2: Training and validation MAE curves during the training of the Allegro MLIP. The E/atom MAE finished around 60 meV/atom and the force MAE finished around 4 meV/Å.*

## 4 Validation

To ensure reliability and to further validate the performance of the trained MLIP, we perform two tests. In the first test, we calculate the phonon dispersion curves of *h*-BN using MLIP and compare them to those obtained by DFT, to evaluate how well MLIP is suitable for describing the lattice dynamics of pristine *h*-BN. In the second test, we compare defect structures relaxed with the MLIP and DFT to ensure the good structural predictions of the MLIP.

### 4.1 Phonon Dispersion

To compute the dispersion of phonons with MLIP, we first optimized the lattice constant for pristine *h*-BN using MLIP and utilized Allegro's capability to interface with ASE.[37] Phonon dispersion curves were then calculated with ASE's Phonon calculator, using the finite different method. We used a 10×10×1 supercell for obtaining converged phonon dispersion. We used the finite difference method as embodied in the code Phonopy[38,39] to compute phonon dispersion based on input from DFT for which we used the ab initio electronic structure code VASP[30,40]. Our computational supercell also consisted of a 10×10×1 cell of single layer *h*-BN. Figure 3 shows a direct comparison between the phonon dispersion for pristine single layer *h*-BN obtained using our MLIP and that obtained using DFT. We see that the MLIP predicts the dispersion of the acoustic modes in good agreement with those computed by DFT, only deviating very slightly. Likewise the optical modes are also in good agreement with those computed by DFT, but there is a deviation from DFT's prediction values. This is actually consistent with what we see from other MLIPs in literature. For example Thiemann et al [41] and Qian and Yang [42] find a similar deviation using a Gaussian Approximation Potentials, i.e., seeing a high level of accuracy in the acoustic modes and a slight shift in the optical modes. This agreement shows that our MLIP is accurate for pristine systems of *h*-BN while retaining the speed benefit that is desired of MLIPs. It is important to note that despite this small deviation the MLIP predicts the optical modes far better than an empirical potential. In Mandelli et al. [43] we can see that the Tersoff potential agrees well with DFT and experimental results in the acoustic modes, but in the optical modes it fails to predict the transverse and longitudinal modes accurately, with errors of >30meV in some places when comparing to DFT or experiment.

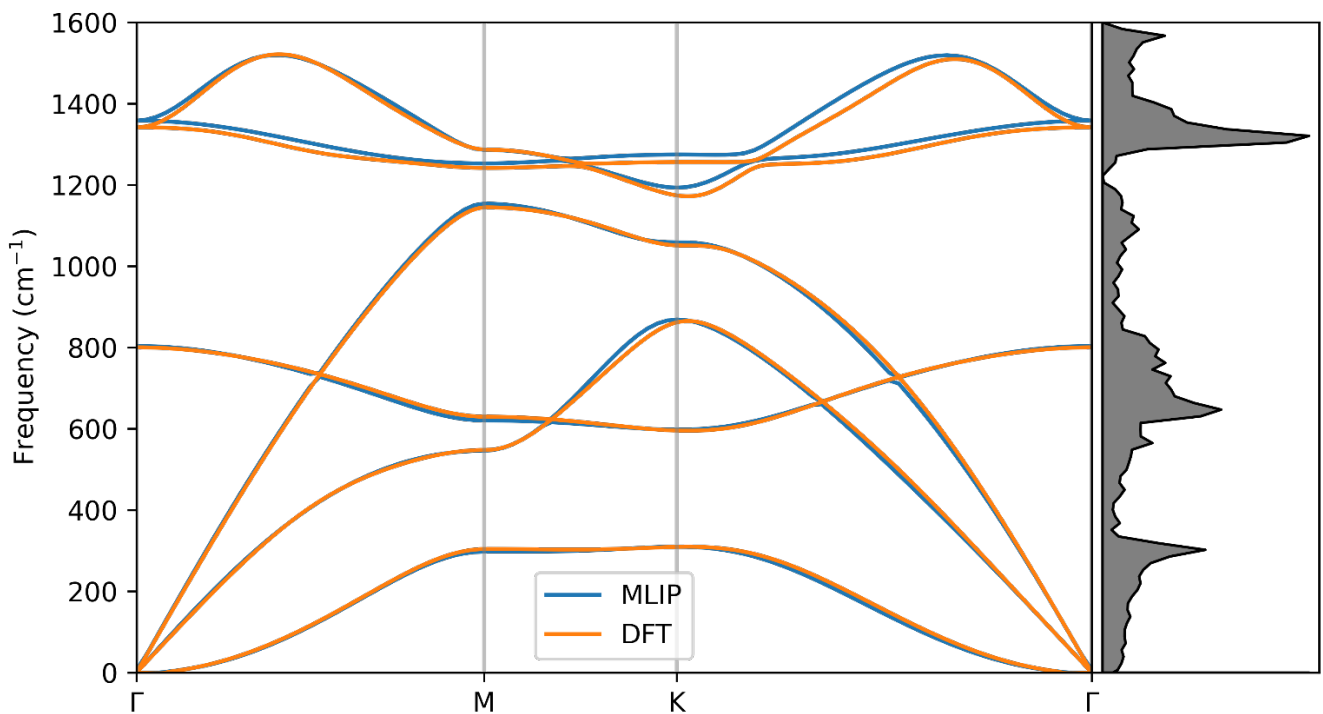

*Figure 3: Phonon dispersion curves of pristine h-BN calculated with DFT (orange) and our MLP (blue). The vibrational density of states is plotted on the right.*

## 4.2 Defect Structures

To evaluate the ability to predict geometric structures of defect-laden $h$-BN using our MLIP, we carried our ionic relaxation of single layer $h$-BN with each of the 6 point defects, shown in Figure 1, with both DFT and the MLIP and compared the resulting bond lengths and angles. We used the same force threshold of 0.02 eV/ Å for calculation with DFT and MLIP. In addition to the point defects, we also relaxed and compared the structure for a grain boundary to ensure that the MLIP correctly simulates defects beyond simple point defects. Figure 4 shows the parity plot of interatomic distances between every pair of atoms for each structure that was relaxed by both DFT and the MLIP. We found an excellent agreement between MLIP and DFT for not only the local environment around each atom, but also for the extended environment up to 20 Å. The mean absolute error (MAE) of all the interatomic distances is 0.024 Å, a small value suggesting that our MLIP correctly handles the extended environment around these defects beyond the cutoff and into the nanometer range.

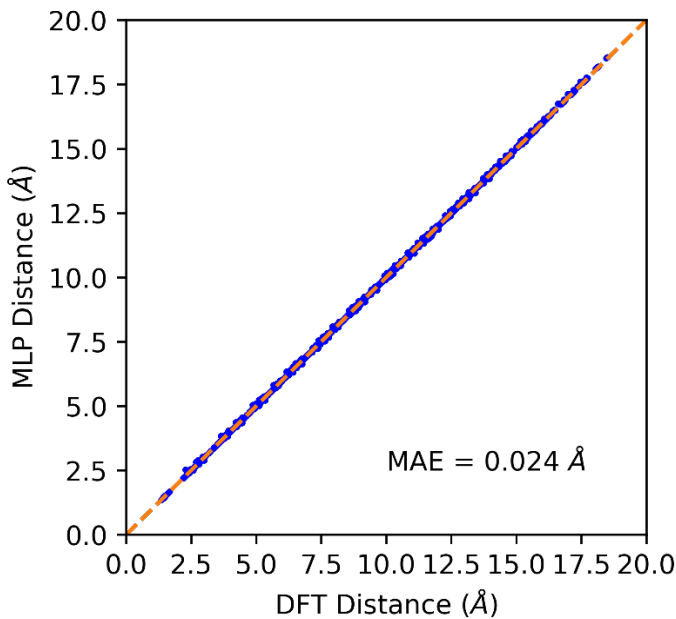

*Figure 4: Direct comparison between MLIP and DFT interatomic distances in the relaxed defect structures.*

In order to obtain a more detailed comparison of the local structure near defects and to assess how well our MLIP performs relative to benchmark data by DFT, we analyzed the bond lengths and bond angles surrounding the defect sites. Table 1 summarizes this comparison across the six different defect structures shown in Figure 1, revealing that the bond lengths and angles obtained through our method are remarkably similar to those calculated using DFT. The agreement between our MLIP and DFT demonstrates the reliability of our MLIP for capturing local geometric distortions around defects. The largest discrepancy occurs in the bond lengths found near the boron vacancy, where the maximum error reaches 0.13 Å. This deviation remains within acceptable limits for the accuracy of MLIP. Similar excellent agreement between our MLIP and DFT was also found for the Stone-Wales defect structure, as shown in Figure S.4 in the SI.

**Table 1**: *Bond length and angle comparisons between DFT and MLP relaxed structures for the 180, BN, NB, VB, VN, and 558 grain boundary defects.*

|   | B/N DFT | B/N MLIP | $B_N$ DFT | $B_N$ MLIP | $N_B$ DFT | $N_B$ MLIP | $V_B$ DFT | $V_B$ MLIP | $V_N$ DFT | $V_N$ MLIP | 558 GB DFT | 558 GB MLIP |
|---|---|---|---|---|---|---|---|---|---|---|---|---|
| **Comparing bond length (Å)** | | | | | | | | | | | | |
| A | 1.483 | 1.484 | 1.452 | 1.452 | 1.474 | 1.474 | 2.288 | 2.253 | 2.685 | 2.823 | 1.458 | 1.444 |
| B | 1.428 | 1.430 | 1.418 | 1.432 | 1.438 | 1.440 | 2.287 | 2.253 | 2.685 | 2.823 | 1.656 | 1.706 |
| C | 1.379 | 1.378 | 1.418 | 1.432 | 1.597 | 1.602 | 1.463 | 1.436 | 1.407 | 1.397 | 1.457 | 1.444 |
| D | 1.599 | 1.601 | 1.452 | 1.452 | 1.597 | 1.602 | 1.438 | 1.451 | 1.441 | 1.459 | 1.442 | 1.444 |
| E | 1.431 | 1.432 | 1.444 | 1.444 | 1.438 | 1.440 | | | | | 1.471 | 1.473 |
| F | | | | | | | | | | | 1.444 | 1.432 |
| **Comparing bond angle (°)** | | | | | | | | | | | | |
| 1 | 117.13 | 117.19 | 120.40 | 120.68 | 117.02 | 116.84 | 125.70 | 125.41 | 114.87 | 118.94 | 104.34 | 106.36 |
| 2 | 124.20 | 124.10 | 120.39 | 120.68 | 125.22 | 125.16 | 117.58 | 120.90 | 119.89 | 121.18 | 117.13 | 115.27 |
| 3 | | | | | | | | | | | 107.10 | 106.00 |

To demonstrate that our MLIP is capable of simulating structure of large-scale defects, we compare the structure of a 558 grain boundary relaxed with our MLIP and with DFT. It is important to note that the structure of this grain boundary or of any other grain boundary is not part of our training dataset. Despite that fact, our MLIP predicts the geometric structure that is in great agreement with that predicted by DFT, as summarized in Table 1. The largest difference that we see between the MLIP and DFT structures is that of 0.05Å. The other differences that that we see are all around 0.01Å or even less, showing the high agreement between DFT and the MLP relaxed structures. This demonstrates the ability of our MLIP to simulate the system beyond those included in training dataset.

## 5  Grain Boundary Motion

Grain boundaries in *h*-BN present both opportunities and challenges for material scientists. These structural defects, which inevitably form during *h*-BN growth, prevent the material from achieving single-crystal quality and can exhibit fascinating properties that may enable unique applications. However, for many high-performance electronic and optoelectronic applications that require pristine material characteristics, grain boundaries are detrimental to performance. Thus, developing effective strategies to eliminate grain boundaries has become a critical research priority. One particularly promising approach involves the controlled manipulation of grain boundary dynamics, which could lead to systematically directing grain boundaries to migrate toward the material's perimeter through controlled motion, effectively expelling these defects from the h-BN sheet. This grain boundary engineering technique could potentially yield large-area, single-domain h-BN with superior crystalline quality. These grain boundary motions have been investigated experimentally [24]. It was shown that 4|8 grain boundaries in monolayer $h$-BN move along the B-N pairs in the lattice via a bond rotation of single-row atoms. This movement results in a diagonal movement of the grain boundary units by a distance of $\sqrt{3}/2\ a$ where $a$ is the lattice constant for the BN. This movement is repeated to give the full gliding motion of the grain boundary over a distance of $\sqrt{3}\ a$ perpendicular to the direction of the grain boundary. The motion viewed experimentally suggests a predictable motion of the grain boundary, but to be able to fully explain the grain boundary motion we must start by investigating the mechanism behind the movement of the grain boundary.

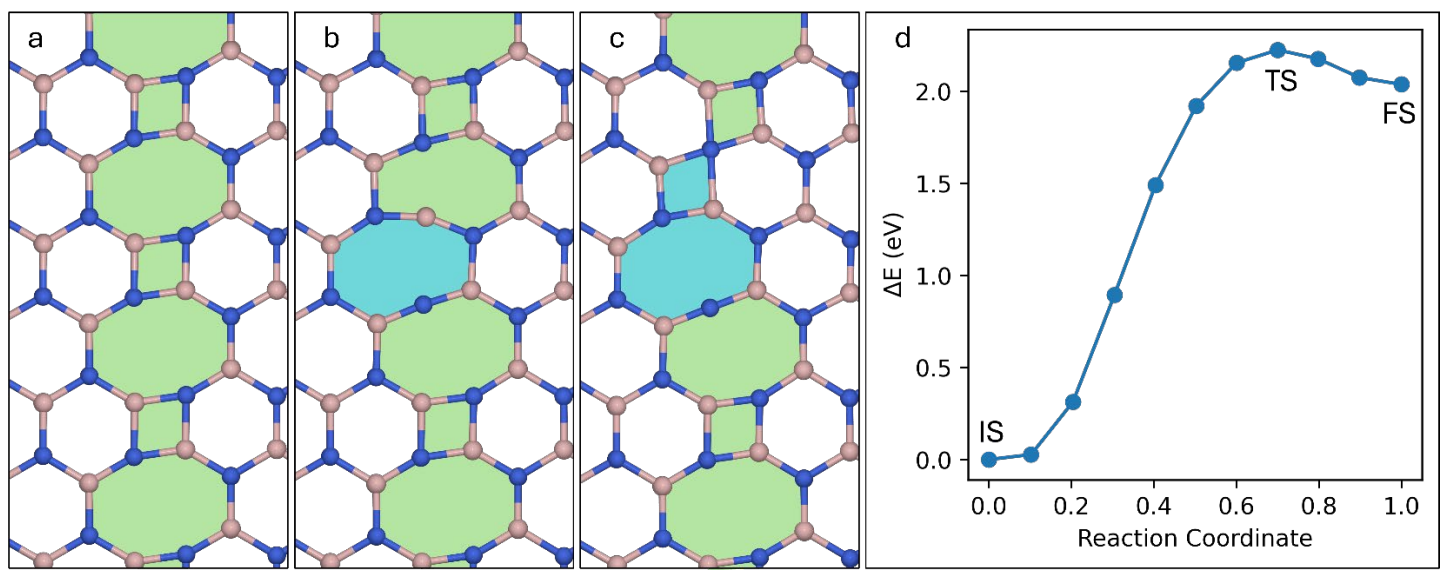

Figure 5: a-c) Illustration of movement of a 4|8 grain boundary. A boron atom in the grain boundary moves, creating a new grain boundary unit. d) Energy barrier for the movement of the first grain boundary unit obtained using NEB calculations with the MLP. The initial, transition, and final states labeled in D have their structures shown in A-C respectively.

In this work, we use our robust MLIP to study this experimentally observed motion of the 4|8 grain boundary (Figure 5a). We first compute the activation barrier, using the NEB method, for moving one B atom up, i.e., creating the starting motion. The movement of a 4|8 grain boundary is done almost exclusively by moving boron atoms in the grain boundary. By moving a boron atom in one of the 4-atom units, a new grain boundary unit is formed, effectively moving the grain boundary in a direction perpendicular to the direction of the grain boundary. The IS, TS and FS are shown in Figure 5a, b, c. In Fig 5b we see the transition state for the grain boundary movement where the moving boron atom breaks its bond with the nitrogen in the 4-atom unit and moves towards its new 4-atom unit. The energy profile along the reaction pathway is shown in Figure 5d. The energy barrier for the movement of the first grain boundary unit, shown in Figure 5d, is large (~2.2 eV), implying that the grain boundary has very little chance of moving naturally at any temperature below the melting point of BN. Performing NEB calculations for unit movements after the first unit shows that the barrier drops dramatically, to less than 0.5 eV indicating that the later unit movements are much easier and more likely to occur. Figure 6 shows the forward and reverse energy barriers for the first 10 movements and shows that both the forward and reverse barriers become nearly equivalent starting at the movement of the 4$^{th}$ grain boundary unit.

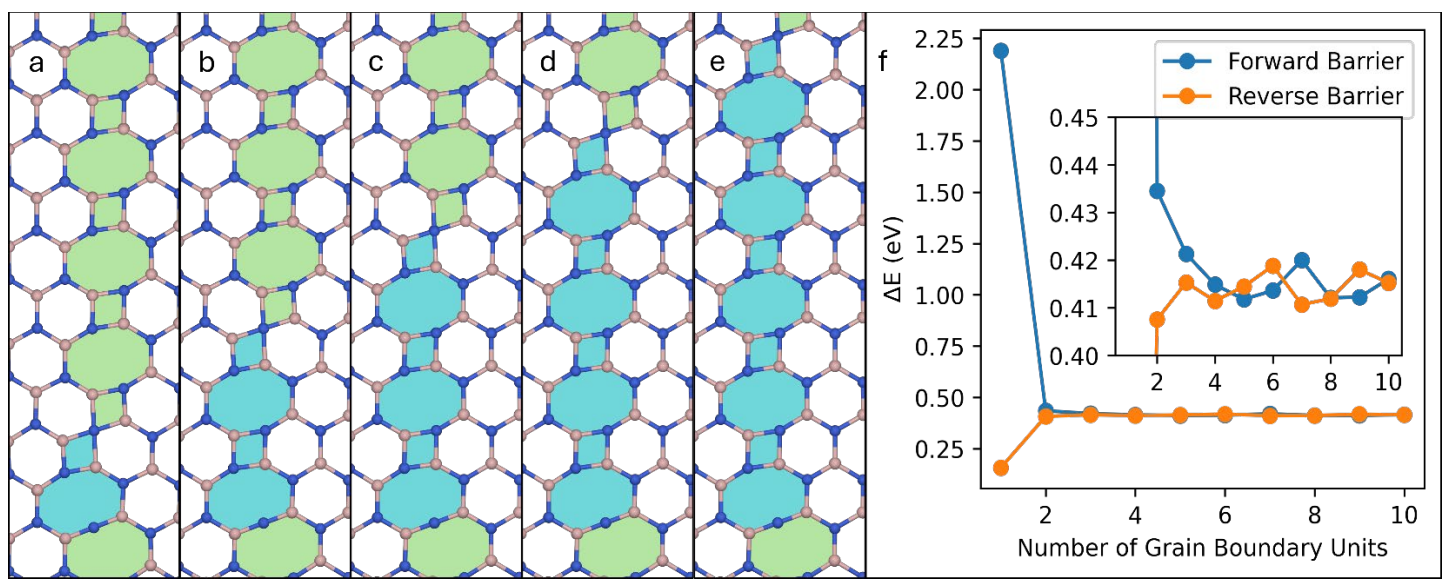

Figure 6: a-e) Atomic structures of 4|8 grain boundary after the first 5 grain boundary unit movements. f) Forward and reverse energy barriers for the first 10 grain boundary unit movements. Inset shows the details of the barrier changes for the 2$^{nd}$ and farther units.

The final aspect of 4|8 grain boundary motion we investigated using the MLP was to perform MD simulations at different temperatures to get the average grain boundary position over time. We performed 12 separate

simulations of the grain boundary under an NVT ensemble for temperatures of 500°K, 750°K, 1000°K, 1250°K, 1500°K, 1750°K, and 2000°K (resulting in a total of 84 simulations). Each simulation started from the same atomic structure containing 12,060 atoms but each simulation had different initial velocities. The simulations ran for 500 ps, and all the data fitting ignored the first 10ps from each simulation to allow the system to come to a thermal equilibrium before data was considered significant. Due to the high energy barrier for the first unit movement, as shown in Figure 9, we constrained the first grain boundary unit to be moved and stay moved as the small reverse barrier would have caused the grain boundary to return to normal instead of continuing to move. All other atoms were allowed to move freely, but the boron atom involved in the first unit movement was constrained to be a certain distance from a nitrogen atom in the final moved configuration. Even with this constraint we still found that below 1000K the grain boundary didn't move, but at temperatures above 1250K we saw that the grain boundary had significant movement, with the probability for a full movement of the grain boundary increasing with increasing temperature.

Following the interface random walk method detailed in [44] and adapting it to be applicable to a 1D interface, we take the grain boundary position to be $h(x,t)$ where the position is limited to the $x$ direction since the motion of the boundary is primarily in the $x$ direction. The velocity of the interface is given as $v = M\rho$ where $M$ is the mobility of the interface and $\rho$ is a driving force which is dependent on intrinsic thermal fluctuations as well as interface curvature. Using the small slope approximation and taking our interface movement to be in the $x$ direction we can express the velocity as a function of the interface position as shown in Equation 1, where $\Gamma$ is the interface stiffness and $\eta$ is the contribution from thermal fluctuations.

$$v = \frac{dh}{dt} = M(\Gamma * \frac{d^2h}{dx^2} + \eta) \qquad (1)$$

Once we integrate both sides in space and time we get Equation 2 where $\bar{\eta}$ is the average force contribution from thermal fluctuations. Taking the variance of Equation 2 we get Equation 3, which indicates that the interface will perform a classical random walk such that the variance of the average interface position increases linearly with time.

$$\bar{h} = M \int_0^t \bar{\eta}\, dt' \quad (2)$$

$$<\bar{h}^2> = \frac{2Mk_BT}{L} t \quad (3)$$

Computationally, we track the average grain boundary position over time and take the variance of the positions of simulations at the same temperature. To simplify slightly, and to allow for easier comparison with experimental results [24], we plot $<\bar{h}^2> * L$ where L is the length of the grain boundary. From the plots shown in Figure 7, the slope of the linear fits relates to the mobility, $M$, through Equation 4, where $s$ is the slope of the linear fit and $T$ is the temperature of the simulation.

$$M = \frac{s}{2k_BT} \quad (4)$$

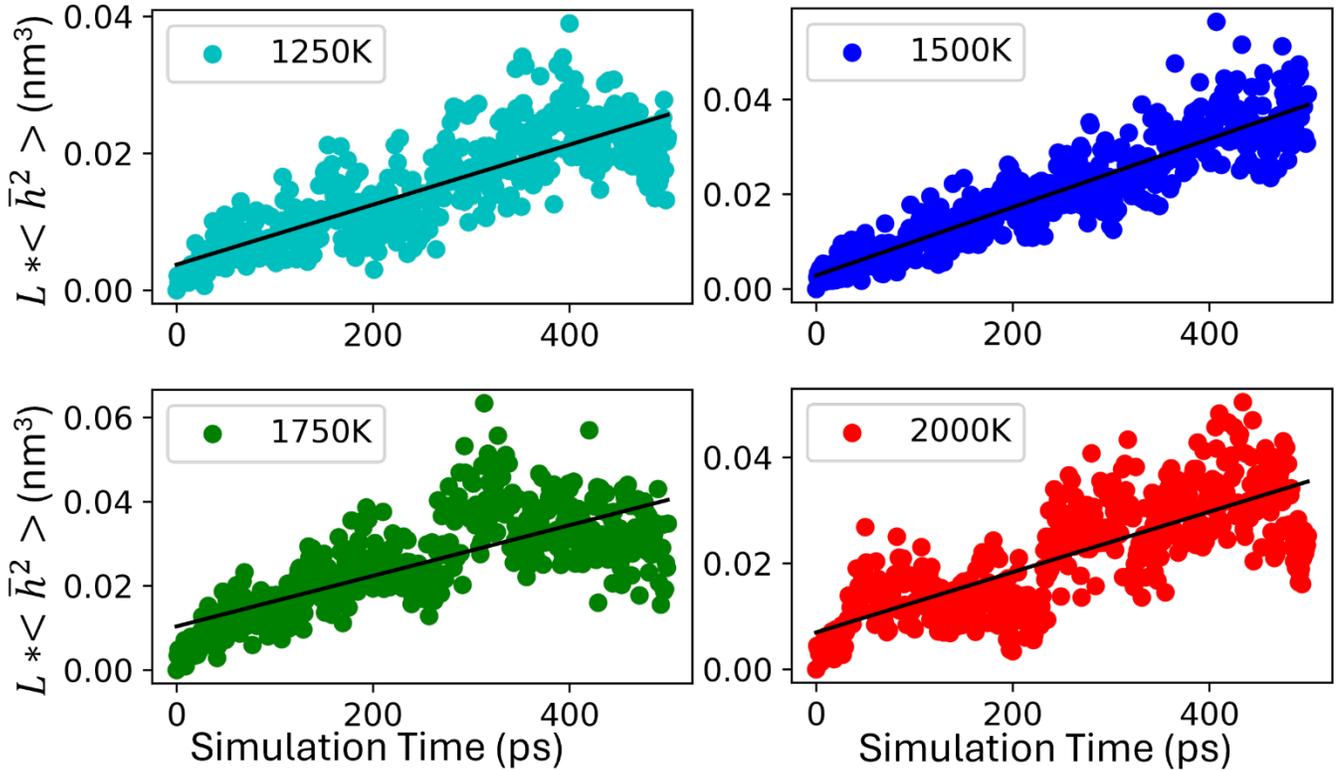

Figure 7: Grain boundary length multiplied by variance of the average grain boundary position ($L * <\bar{h}^2>$) for the higher temperatures used in simulations along with their respective linear fits.

Our calculations give a mobility of $1.74 \frac{m^3}{Js}$ at 1500K. To be able to compare to experimental values we need to use the method for computing the mobility scale as described in Ref. [24,44,45]. We do this by using Equation 5, which comes from a unit analysis of Equation 4, for a characteristic time scale ($t_c$) of 10ps and a characteristic length scale ($l_c$) of 1Å to get an scaling value of $\sim 4.8 \frac{m^3}{Js}$. We can compare this scaling value to that from experimental characteristic time and length scales of 1s and 1Å [39] and see the experimental scaling is $\sim 4.8 * 10^{-11} \frac{m^3}{Js}$.

$$M_{Scale} \approx \frac{l_c^3}{k_B T t_c} \tag{5}$$

For comparison of our computational results with those extracted from experiments we thus need to apply a scaling factor of $\sim 10^{-11}$. With this scaling factor our computational value becomes $1.74 \times 10^{-11} \frac{m^3}{Js}$ which is slightly smaller than the experimental value of $1.36 * 10^{-9} \frac{m^3}{Js}$. Note, however, that in the experiments energetic e-beam irradiation was used as a stimulus for the grain boundary motion [24] which would lead to a higher measured mobility than is obtained in simulations. Our obtained value for the mobility of the grain boundary is thus reasonable.

The above comparison of our computational findings to the experimentally viewed grain boundary motion shows that the MLIP that we have developed is reliable for simulations with tens of thousands of atoms, robust enough to accurately handle complex and large-scale defects such as grain boundaries, and maintains DFT level accuracy while being orders of magnitude less computationally expensive. Putting all those attributes together, our MLIP could be applied to a wide range of problems including geometric optimization of large defects or collections of defects which would be useful for helping to identify chemically active sites or the evolution in time of these larger defects whose dynamics are difficult to access through other means.

# 6 Conclusion

We have developed a machine learned potential, using the Allegro framework, for hexagonal boron nitride with and without defects that shows near DFT-level accuracy. We tested the performance of the potential against DFT using phonon dispersions, relaxed defect structures, and grain boundary formation energy, and for each metric, the potential reproduced the DFT results with minimal errors. We also explored the movement of grain boundaries and uncovered the details of the movement, showing that if the first grain boundary unit movement can be stimulated, by an external stimulant such as electron beam irradiation, the rest of the grain boundary will follow suit, allowing the effective removal of grain boundaries and extending the area of pristine geometry. While the potential shown is successful in imitating a potential for $h$-BN to DFT-level accuracy, we want to emphasize the importance of some of the lessons that we learned and the pitfalls we encountered. By adding higher energy or metastable structures to the training data, the potential learns not only what is correct (i.e. structures, energies, and forces near equilibrium) but it also learns what is "wrong" (i.e. structures, energies, and forces not near an equilibrium). This expands the accessible regime of the potential and makes it less likely for the potential to incorrectly extrapolate due to a given structure being outside that accessible regime.

**ACKNOWLEDGMENTS**

This material is based upon work supported by the U.S. Department of Energy, Office of Science, Office of Basic Energy Sciences Catalysis Science program under Award Number DE-SC0024083. This research used resources of the National Energy Research Scientific Computing Center (NERSC), a Department of Energy User Facility using NERSC award BSE-ERCAP0033715.

Thanks to the input from Dr. Miguel Caro, we learned the importance of having "incorrect" structures in the training dataset.

# Supplementary information for: Machine learned potential for defected single layer hexagonal boron nitride

John Janisch,[1] Duy Le,[1,2] and T.S. Rahman[1,2,3]

[1]Department of Physics, University of Central Florida, Orlando, Florida 32816, United States

[2]Renewable Energy and Chemical Transformations (REACT) Cluster, University of Central Florida, Orlando, Florida 32816, United States

3Donostia International Physics Center, Paseo de Manuel Lardizábal 4, 20018, San Sebastian, Spain.

**Attempt at MLIP for *h*-BN with Behler and Parrinello artificial neural network potential**

We used the PROPhet code[1] to train the MLIP, employing the Behler and Parrinello[2-4] artificial neural network approach (ANN). We trained and tested 10 different ANN by varying the neural network architecture, the number of gaussian descriptors, and the radial cutoff. We find that a neural network with two hidden layers of 48 nodes, 24 radical and 24 angular gaussian descriptors with 6Å cutoff radius is the most accurate one.

The ANN potential was capable of reproducing the general structure of *h*-BN, but the relaxed atomic structure of a grain boundary displayed discrepancies around the grain boundaries when comparing to the structure obtained using DFT codes such as VASP [5,6]. As shown in Figure S1, the areas of pristine *h*-BN do not match exactly the DFT results, but they do hold up the general hexagonal structure and show the same lattice constant. When we shift our focus to the grain boundaries the story is quite different, as there seems to be no rationale for the relaxed atomic positions obtained near the grain boundary by the ANN potential.

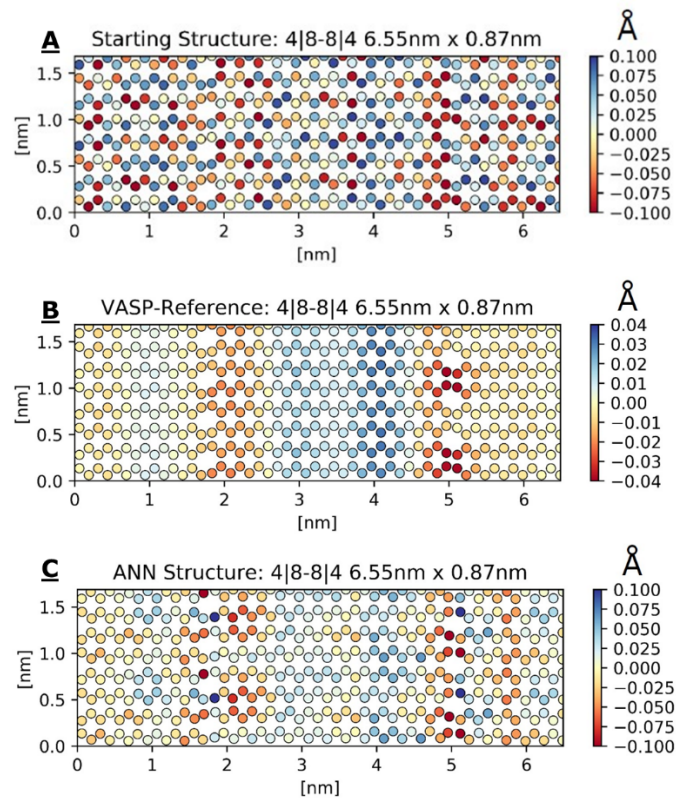

*Figure S1: A) Initial atomic configuration used for relaxations by DFT and ANN potentials. B) Relaxed structure using DFT (VASP). C) Relaxed structure with ANN. We see the ANN potential gets the same general structure, but the area near the grain boundary has larger discrepancies.*

Increasing the size of the simulation highlights a much larger difference between ANN and DFT results. As shown in Figure S2, the local area around the grain boundaries varies quite significantly from the structure relaxed with DFT, but areas farther from the grain boundaries are quite similar. A plot of the difference in the z-coordinates between the structures obtained by DFT and ANN, shown in Figure S2, reveals that a majority of the errors are grouped around the grain boundaries.

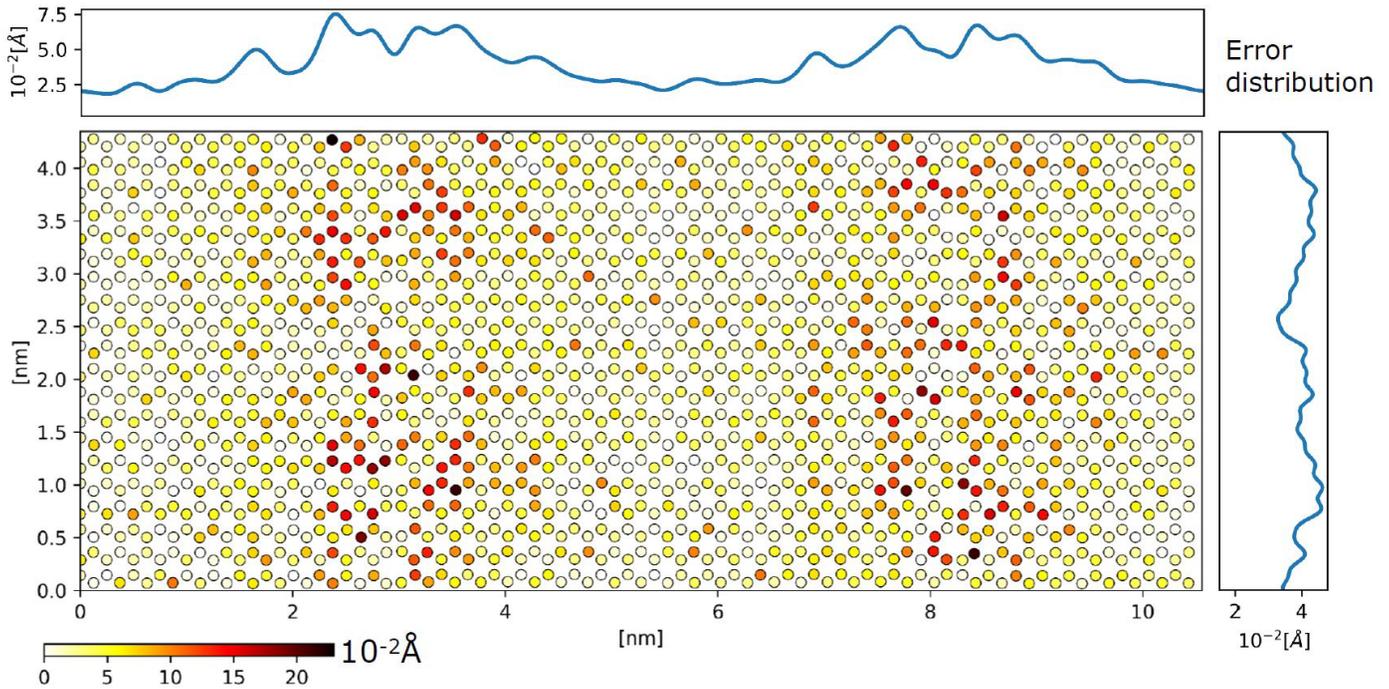

*Figure S2: Difference in z position of atoms between DFT and ANN relaxed structures. The errors are localized around the grain boundaries, which indicates that the ANN potential is having issues with the long-range effects of the grain boundaries.*

Figure S3 shows the grain boundary formation energy computed with ANN and DFT. We found that the ANN failed to capture the effect of distance between the grain boundary on the grain boundary formation energy. While the ANN potential (Fig. S3B) shows an improved trend in the grain boundary energy, all the ANN potentials had an issue.

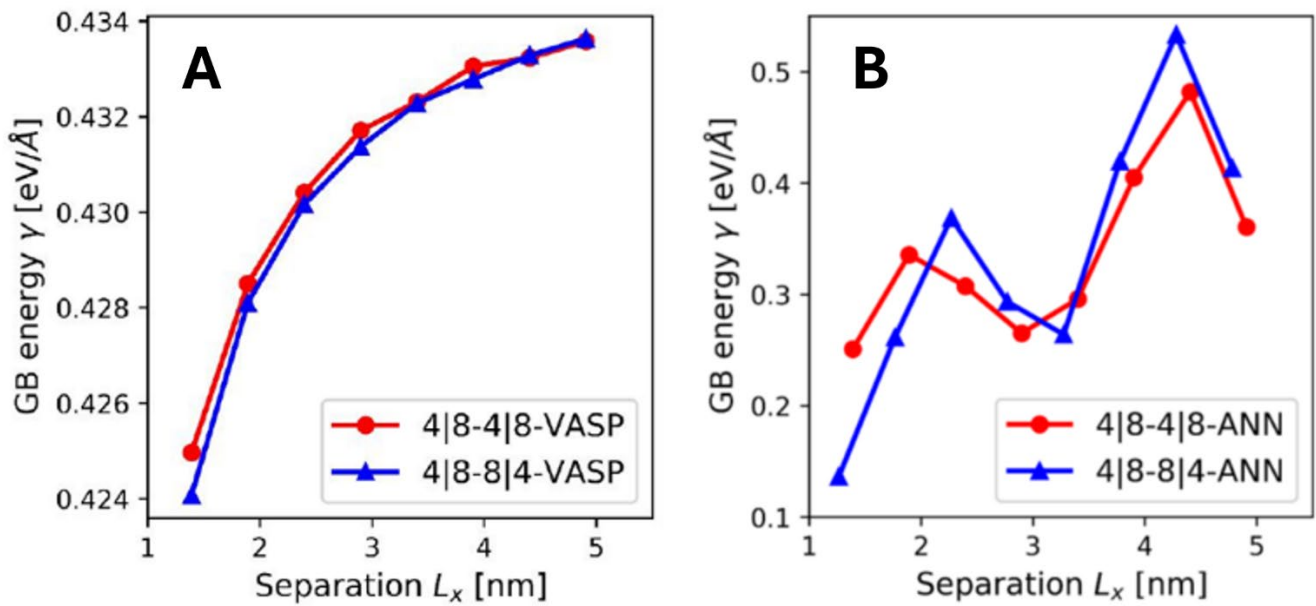

*Figure S3: Comparison of grain boundary formation energies obtained with DFT codes VASP (A) and ANN potentials*

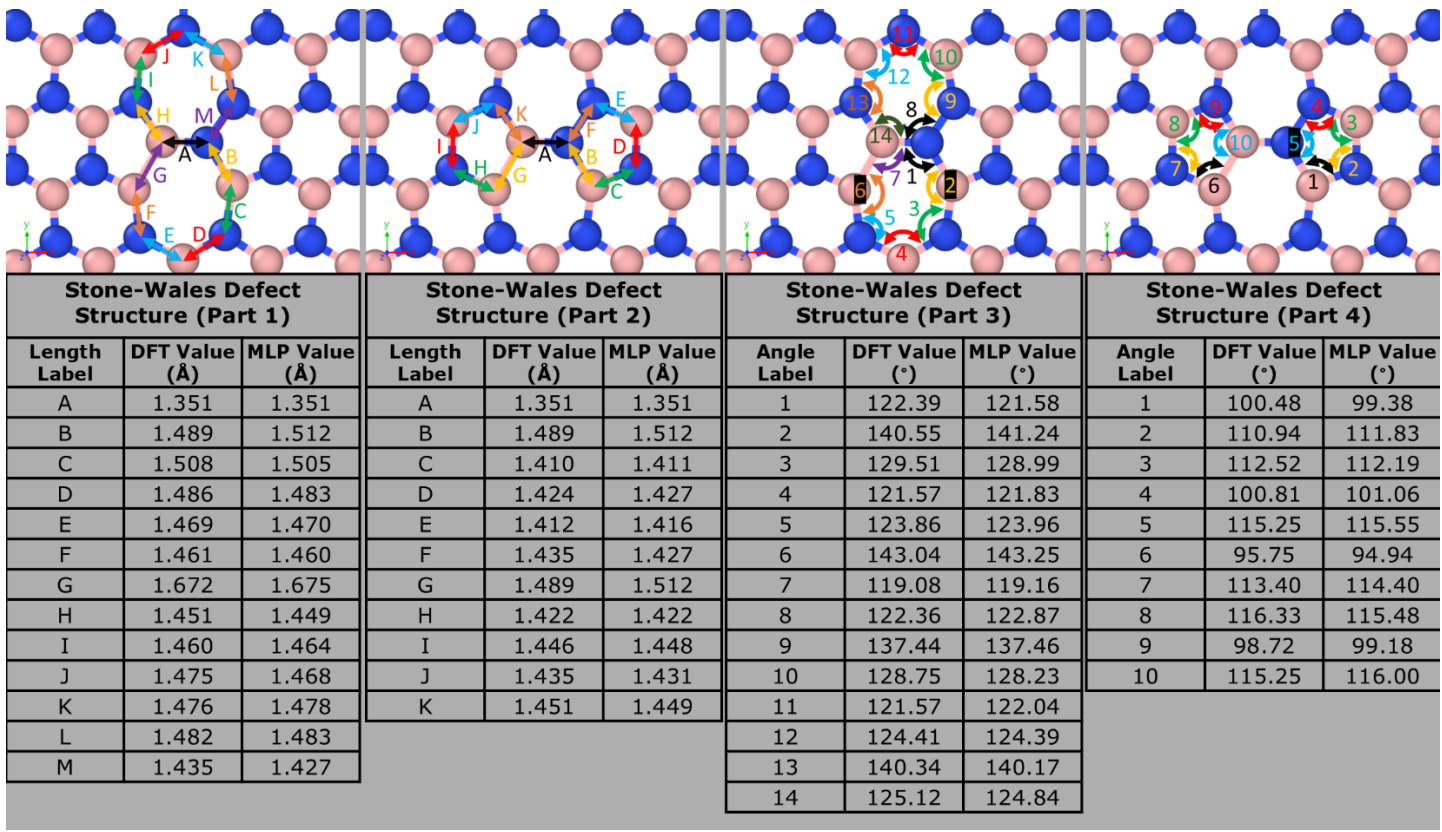

*Figure S4: Bond lengths and angles for Stone-Wales Defect. Through the large number of lengths and angles, we see that the agreement between DFT and the ANN is very good, showing that even for larger defects the ANN accurately recreates the forces that atoms are experiencing.*

Another small defect we looked at was the Stone-Wales defect in which a BN pair is rotated by 90 degrees, creating a defect that has more directly affected atoms than a vacancy or atom replacement, and therefore has more bond lengths and angles that are worth comparing. Figure S5 shows all the bond lengths and angles that we compared in detail between DFT and the Allegro MLP. We still see the same trend we saw with all the other defects, that being that there is little difference between DFT and the MLP when it comes to the relaxed bond lengths and angles.

From the above we conclude that the ANN potential was not capable of producing correct forces for the grain boundary regions, and that long-range effects were not captured by the ANN potential. Moreover, molecular dynamics (MD) simulation with the ANN potential always resulted in the simulation crashing, typically within a few picoseconds. Without the ability for an ANN potential to produce a stable MD simulation we decided to change the machine learned model to Allegro.